# Ions in glass forming glycerol: Close correlation of $\alpha$ and fast $\beta$ relaxation

M. Köhler,[1] P. Lunkenheimer,[1,*] Y. Goncharov,[2] and A. Loidl[1]

[1]*Experimental Physics V, Center for Electronic Correlations and Magnetism, University of Augsburg, 86135 Augsburg, Germany*
[2]*Institute of General Physics, Russian Academy of Sciences, 119991 Moscow, Russia*

We provide broadband dielectric loss spectra of glass-forming glycerol with varying additions of LiCl. The measurements covering frequencies up to 10 THz extend well into the region of the fast $\beta$ process, commonly ascribed to caged molecule dynamics. Aside of the known variation of the structural $\alpha$ relaxation time and a modification of the excess wing with ion content, we also find a clear influence on the shallow loss minimum arising from the fast $\beta$ relaxation. Within the framework of mode-coupling theory, the detected significant broadening of this minimum is in reasonable accord with the found variation of the $\alpha$-relaxation dynamics. A correlation between $\alpha$-relaxation rate and minimum position holds for all ion concentrations and temperatures, even below the critical temperature defined by mode-coupling theory.

In broadband dielectric loss spectra of glass forming materials, a number of different dynamic processes is revealed leading to a conglomerate of spectral features like peaks, power laws, or minima that arise in the whole accessible frequency range from µHz to THz [1,2,3]. Except for the so-called $\alpha$ peak, which is mirroring the structural relaxation processes that drive the glassy freezing at the glass transition temperature $T_g$, the microscopic origin of the other spectral features is still controversially discussed. Elucidating the nature of these processes seems to be a prerequisite for achieving a deeper insight into the mechanisms of the glass transition and the glassy state of matter in general. Especially, it seems clear nowadays that the different spectral features are closely interrelated and cannot be described independently.

In the present work, we examine those interrelations by providing broadband dielectric spectra of one of the most studied glass formers, glycerol, whose dielectric response is being systematically modified by the addition of ions. It is well known that dissolving salts like, e.g., LiCl in glycerol leads to strong shifts of the $\alpha$-relaxation peaks [4,5]. Thus, the question arises if the behavior of the faster processes like excess wing, fast $\beta$ process, and boson peak, which all are observed in pure glycerol [2,6] is also affected by the ion addition. For example, according to mode coupling theory (MCT), which was shown to describe a variety of dynamic properties of pure glycerol [7,8,9,10], the temperature dependences of such different quantities as the $\alpha$-relaxation rate (located in the mHz to GHz range) and the frequency of the loss minimum showing up in the 10 – 100 GHz regime are predicted to be closely correlated.

Several experimental techniques were combined to arrive at extreme broadband dielectric spectra covering a frequency range of 20 Hz < $\nu$ < 18 THz. This includes a standard LCR-meter technique at $\nu$ < 1 MHz, a coaxial reflection method at 1 MHz < $\nu$ < 2 GHz, a coaxial transmission technique at about 100 MHz < $\nu$ < 40 GHz, a quasi-optical Mach-Zehnder interferometer at 60 GHz < $\nu$ < 1.2 THz, and a Fourier-transform infrared spectrometer at about 1 THz < $\nu$ < 18 THz. For cooling and heating of the samples, a nitrogen gasheating system was used. For further experimental details, the reader is referred to [2,11]. The sample materials were purchased from MERCK and measured without further purification. The specified purity for glycerol was ≥ 99.5%. The LiCl concentrations are given in mol%.

Figure 1 shows the frequency dependence of the dielectric loss $\varepsilon''$ at three selected temperatures and for four LiCl concentrations. The results on pure glycerol were taken from [3,12]. The behavior of the $\alpha$ relaxation in dependence of LiCl admixture was treated in detail in [5] and is only briefly discussed in the following. With increasing ion content, the $\alpha$ peaks become superimposed by the conductivity contribution of the ionic charge transport, which in the loss spectra shows up as a $1/\nu$ divergence at low frequencies. As directly seen in Fig. 1 and corroborated by an analysis including the conductivity contribution [5], the $\alpha$ peaks shift towards lower frequencies, i.e., the structural $\alpha$-relaxation times $\tau_\alpha$ increase with increasing salt content. This can be ascribed to a reduced molecular mobility caused by interactions between the glycerol molecules and ions.

Another effect of ion addition, which was not seen in the spectra restricted to the $\alpha$-peak region that were provided in [5], is the variation of the so-called excess wing at low temperatures (e.g., 213 K in Fig. 1). This second, more shallow power law at the high-frequency flank of the $\alpha$ peak shows a marked reduction of slope, i.e. of its power-law exponent, for increasing salt concentration. Moreover, it becomes more pronounced compared to the high-frequency flank of the $\alpha$ peak. It was shown that the excess wing can be ascribed to a secondary relaxation process leading to a loss peak that is partly submerged under the dominating $\alpha$ peak [13,14,15]. Interestingly, recently it was found that the excess wing in glycerol develops into a strong secondary relaxation peak under high pressure [16]. This finding was ascribed to a breakdown of the hydrogen bonds between molecules under high pressure. It seems reasonable that ion addition also can lead to a partial breaking of hydrogen bonds, e.g., via the attachment of Li ions to the oxygen atoms, which would explain the stronger relative strength of the excess wing, observed here.

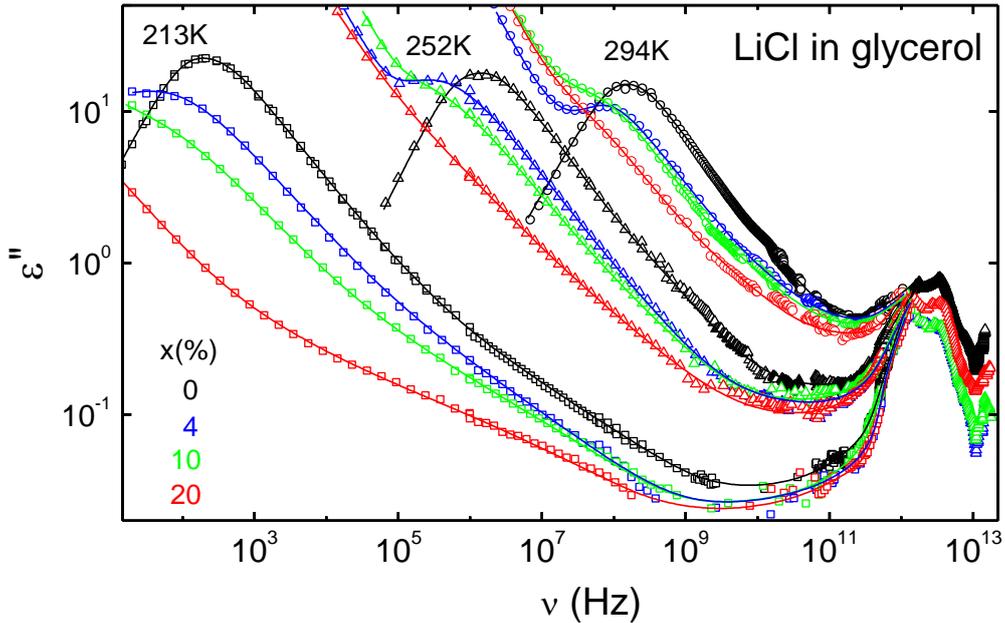

FIG. 1. Comparison of broadband dielectric loss spectra of glycerol-salt mixtures with various LiCl concentrations, shown for three typical temperatures. For better readability, in the boson-peak region ($\nu > 1$ THz) the results are shown for 252 K, only. The lines are guides to the eyes. The data for pure glycerol were previously published in [2,3,7,12] and were taken at 213, 253, and 295 K.

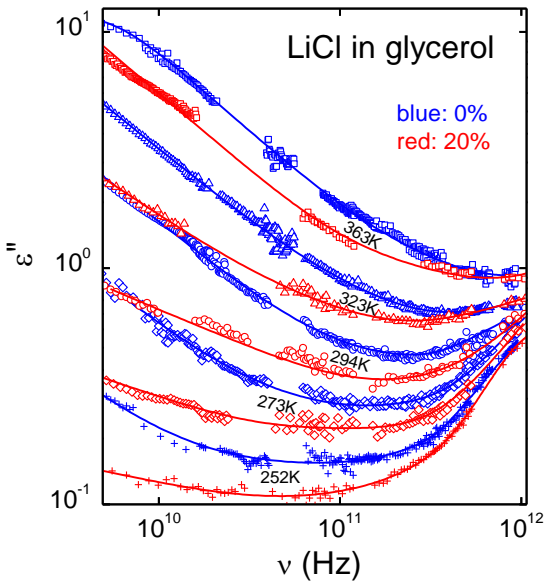

FIG. 2. Dielectric loss spectra in the minimum region of pure glycerol and of glycerol with 20% LiCl, shown for various temperatures. Same temperatures are denoted by identical symbols. The lines are guides to the eyes.

The present broadband spectra, including results in the microwave and millimeter-wave range, reveal that ion addition also strongly affects the minimum region at about 10 – 100 GHz. Here a fast process contributes to the loss of glass-forming liquids, which is often ascribed to the rattling motion of the molecules within the cage formed by its next neighbors [17,18]. As revealed by Fig. 1 and shown for two concentrations in more detail in Fig. 2, this minimum becomes markedly broadened with increasing ion content. In addition, a significant shift towards lower frequencies and a reduction of its absolute value is observed.

Figure 3 shows the minimum region for all four LiCl concentrations investigated. As an example, the dash-dotted line in (b) demonstrates that the minimum cannot be described by a trivial superposition of the $\alpha$ and boson peak. Thus, there is evidence for excess intensity in this region. Similar evidence for a fast process contributing in this spectral region were also found in a variety of other glass formers [2,3,7,19,20,21]. One possibility for the description of this excess contribution is provided by the assumption of a constant loss as predicted by the extended coupling model [18]. Indeed, especially for the lower temperatures and higher salt concentrations, $\varepsilon''(\nu)$ is found to be nearly constant over a relatively broad frequency range (see, e.g., the 252 K curve for 10% salt content in Fig. 3). However, in pure glycerol the dielectric loss minimum and the corresponding feature in the susceptibilities determined by scattering methods [8] have often been as-cribed to the fast $\beta$ process predicted by MCT. It is assumed to arise from the cage effect mentioned above. In the present work, we use an approach in the framework of the basic "idealized" MCT, corresponding to that applied in ref. [7] to dielectric data on pure glycerol. Here, the minimum region is fitted by the sum of two power laws:

$$\varepsilon'' = \frac{\varepsilon_{min}}{a+b}\left[a\left(\frac{\nu}{\nu_{min}}\right)^{-b} + b\left(\frac{\nu}{\nu_{min}}\right)^{a}\right] \quad (1)$$

$\nu_{min}$ and $\varepsilon_{min}$ denote frequency and amplitude of the minimum, respectively. The temperature-independent exponents $a$ and $b$ are interrelated via the so-called system parameter $\lambda$ [17]:

$$\lambda = \frac{\Gamma^2(1-a)}{\Gamma(1-2a)} = \frac{\Gamma^2(1+b)}{\Gamma(1+2b)} \quad (2)$$

where $\Gamma$ denotes the Gamma function.

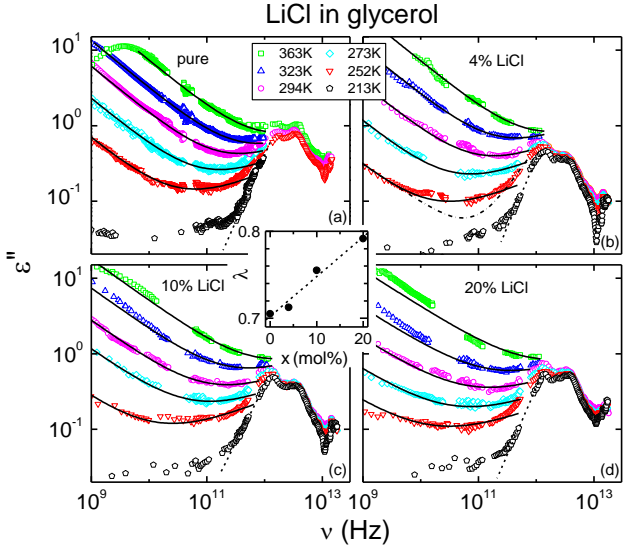

FIG. 3 (color online). Dielectric loss spectra of glycerol-salt mixtures with various LiCl concentrations at frequencies beyond 1 GHz (data for pure glycerol taken from [2,3]). Curves for six different temperatures are shown as indicated in the figure legend. The solid lines are fits with MCT, eqs. (1) and (2). In (b), a curve calculated by the sum of two power laws is shown (dash-dotted line). It demonstrates the presence of additional intensity in the minimum region. The dashed lines indicate the steep increase at the left flank of the boson peak. The inset shows the dependence of the MCT system-parameter on ion concentration. The line in the inset is a linear fit.

The solid lines in Fig. 3 are fits using eqs. (1) and (2), simultaneously applied to all experimental curves at $T \geq 252$ K. (The 213 K curves were not fitted because the number of data points at the minimum is too sparse.) For each concentration, a single system parameter was used to fit the different temperatures. Just as for pure glycerol shown in Fig. 3(a) [2,3,7,12], for the salt solutions the fits provide a good description of the experimental data over 2–3 frequency decades. However, again just as for pure glycerol, at $\nu > \nu_{min}$ only part of the increase towards the boson peak is accounted for by the fits. This can be ascribed to the superposition by the additional, steeper increase towards the boson peak, not covered by MCT [2], which is indicated by the dashed lines in Fig. 3.

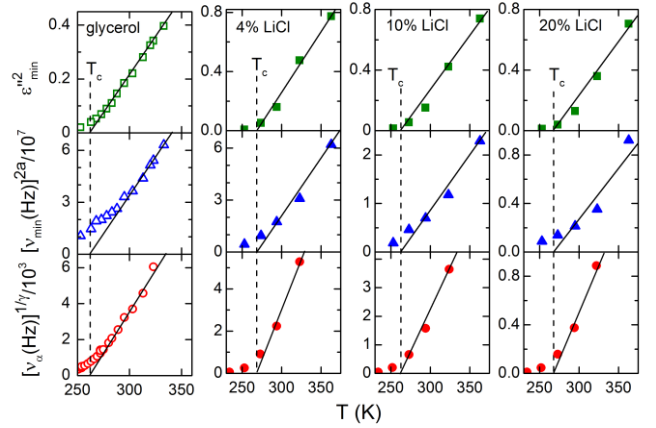

FIG. 4 (color online). Critical temperature dependence of the amplitude (first row) and position (second row) of the $\varepsilon''(\nu)$ minimum and of the $\alpha$-relaxation rate (third row). Data for pure glycerol (first column; [7]) and for glycerol with three concentrations of LiCl are shown. Representations have been chosen that should result in linear behavior according to the predictions of the MCT. The solid lines correspond to the critical laws of MCT and should extrapolate to the critical temperature for ordinate values of zero.

As becomes obvious by the inset of Fig. 3, the system parameter resulting from the fits shows a significant increase with ion concentration. Via eq. (2) this implies a decrease of the exponents $a$ and $b$, i.e. the development of a shallower loss minimum, in agreement with the findings of Fig. 2. Idealized MCT predicts critical laws for the temperature dependences of minimum position and amplitude and of the $\alpha$-relaxation rate, namely $\nu_{min} \propto (T-T_c)^{1/(2a)}$, $\varepsilon_{min} \propto (T-T_c)^{1/2}$, and $\nu_\alpha \propto (T-T_c)^{1/\gamma}$ [with $\gamma = 1/(2a)+1/(2b)$], respectively [17]. Figure 4 provides a check of these laws, which should hold for temperatures above but close to $T_c$ only. If, in addition, taking into account the limited number of data points compared to pure glycerol (left column of Fig. 4), the experimental data can be considered consistent with the theoretical prediction. With values of 262 K (0% LiCl [7]), 268 K (4%), 263 K (10%), and 268 K (20%), Fig. 4 reveals no significant variation of the critical temperature $T_c$ with ion content. It should be noted that the present evaluation within idealized MCT only represents a first check for consistency of the experimental data with MCT and provides a rough estimate of the critical temperature. Only sophisticated evaluations within extended versions of MCT (see, e.g., [9,22]) can reveal more definite information, which, however, is out of the scope of the present work.

Notably, the critical laws of MCT also imply a close correlation of $\alpha$ peak and minimum, i.e., when the temperature dependence of the $\alpha$ peak changes, the shape and position of the minimum also should vary and vice versa. Indeed, this is observed here: The temperature dependence of $\tau_\alpha$ is strongly influenced by ion addition [5] and the minimum broadens and changes its position (Fig. 2). These variations would be quantitatively fully consistent with idealized MCT (at least above $T_c$) if a perfect agreement of fits and experimental data in Fig. 4 could be stated.

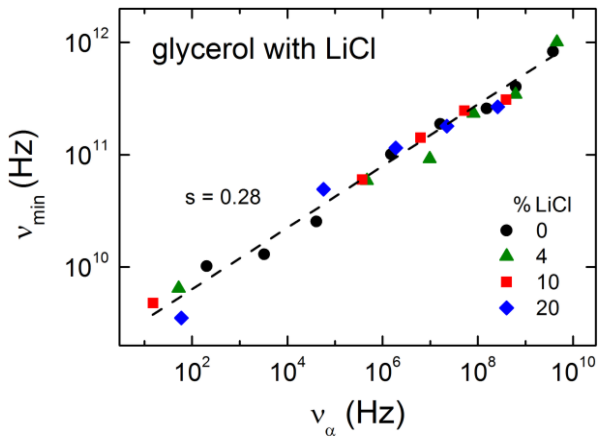

FIG. 5 (color online). Minimum frequency (read off from the experimental spectra) vs. $\alpha$-peak frequencies, estimated by $\nu_p \approx \nu_\alpha = 1/(2\pi\langle\tau_{CD}\rangle)$ with $\langle\tau_{CD}\rangle = \tau_{CD}\beta_{CD}$ where $\tau_{CD}$ and $\beta_{CD}$ were obtained from fits with the Cole-Davidson function [5]. Results are shown for all concentrations and temperatures investigated. The line corresponds to a power law with exponent 0.28.

To further examine these correlations, also including the behavior below $T_c$, in Fig. 5 the experimentally determined minimum and $\alpha$-peak frequencies are plotted against each other. Interestingly, for all investigated temperatures and ion contents a close correlation of both quantities is revealed, which can be approximated by $\nu_{min} \propto \nu_\alpha^{0.28}$. Actually, the critical laws of idealized MCT also lead to a power-law relation of these quantities, namely $\nu_{min} \propto \nu_\alpha^{b/(a+b)}$. For the values of $\lambda$ provided in the inset of Fig. 3, which are related to $a$ and $b$ via eq. (2), we obtain exponents $b/(a+b)$ varying between 0.63 (20% LiCl) and 0.66 (0% LiCl). This is clearly larger than the experimentally observed slope of 0.28. Similar deviations were also found for pure glycerol from light-scattering measurements [23]. However, one should note that a large part of the data points in Fig. 5 were collected below $T_c$ where idealized MCT is invalid and the experimentally observed correlation extends beyond the range expected by idealized MCT. Irrespective of any model considerations, the results of Fig. 5 imply a systematic correlation of the structural $\alpha$ and the fast $\beta$ relaxation despite both processes are separated by many frequency decades. This is not a trivial consequence of the superposition of the $\alpha$ peak (strongly shifting with ion addition [5]) and the boson peak because the minimum clearly arises from a separate process (cf. dash-dotted line in Fig. 3b) and because the excess wing starts to separate $\alpha$ peak and minimum at low temperatures and/or high concentrations (Fig. 1).

In summary, we have provided broadband dielectric spectra of glycerol with varying LiCl salt contents extending well into the frequency region of the fast $\beta$ relaxation and the boson peak. In addition to the well-known slowing down of the $\alpha$ relaxation, ion addition also strongly affects the faster processes: Aside of a significant effect on the excess-wing shape and relative amplitude, most remarkably the minimum marking the fast $\beta$ process becomes significantly broader. According to MCT this should lead to a variation of the temperature dependence of the $\alpha$-relaxation time, which indeed is observed. The findings are qualitatively and partly also quantitatively consistent with idealized MCT but for a thorough check and for understanding the correlations also found below $T_c$, additional analyses within extended versions of MCT or alternative models are necessary. Without considering the details on a molecular level, the present ion addition to glycerol can simply be regarded as the generation of new glass formers, modifying the original system with respect to molecule size (via the attachment of ions to the glycerol molecules) and/or molecule interaction (via the modification of the predominantly hydrogen-bonded network in the pure system). Thus, further systems with similar modifications (e.g., mono-, di-, and trimers [20]) should be studied to check for universal behavior concerning the connection of slow and fast dynamics in glassforming materials.

This work was partly supported by the Deutsche Forschungsgemeinschaft via Research Unit FOR 1394.

———————